\documentclass[prd,
twocolumn,
superscriptaddress,
showpacs,nofootinbib,%
tightenlines]{revtex4-1}
\usepackage{epsfig}
\usepackage{tabularx}
\usepackage{multirow}%
\usepackage{slashed}
\usepackage{extarrows}
\usepackage{amssymb}
\usepackage{xcolor}
\usepackage{color}
\usepackage[colorlinks=true, allcolors=blue]{hyperref}
\usepackage{enumerate}

\newcommand{\be}{\begin{equation}} \newcommand{\ee}{\end{equation}}
\newcommand{\ba}{\left(\begin{array}{c}}
\newcommand{\ea}{\end{array}\right)}
\newcommand{\bea}{\begin{eqnarray}} \newcommand{\eea}{\end{eqnarray}}

\newcommand{\RNum}[1]{\uppercase\expandafter{\romannumeral #1\relax}}

\newcommand{\bma}{\left(\begin{matrix}}
\newcommand{\ema}{\end{matrix}\right)}
\newcommand{\bqa}{\begin{eqnarray}}
\newcommand{\eqa}{\end{eqnarray}}
\newcommand{\bqaa}{\begin{eqnarray*}}
\newcommand{\eqaa}{\end{eqnarray*}}

\allowdisplaybreaks

\begin{document}
\title{Hunting for states in the recent LHCb di-$J/\psi$ invariant mass spectrum}
\author{Ze-Rui~Liang}
\affiliation{School of Physics and Electronics, Hunan University, 410082 Changsha, China}
\author{Xiao-Yi~Wu}
\affiliation{School of Physics and Electronics, Hunan University, 410082 Changsha, China}
\author{De-Liang~Yao}
\email{yaodeliang@hnu.edu.cn}
\affiliation{School of Physics and Electronics, Hunan University, 410082 Changsha, China}
\affiliation{Hunan Provincial Key Laboratory of High-Energy Scale Physics and Applications, \\ Hunan University, 410082 Changsha, China}
\affiliation{School for Theoretical Physics, Hunan University, 410082 Changsha, China}

\begin{abstract}
Partial wave analysis is performed, with effective potentials as dynamical inputs, to scrutinize the recent LHCb data on the di-$J/\psi$ invariant mass spectrum.  Coupled-channel effects are incorporated in the production amplitude via final state interactions. The LHCb data can be well described. A dynamically generated pole structure, which can be identified as the $X(6900)$ state, is found. Our analysis also provides hints for the existence of three other possible states: a bound state $X(6200)$,  a broad resonant state $X(6680)$ and a narrow resonant state $X(7200)$. The $J^{PC}$ quantum numbers of the $X(6680)$ and $X(6900)$ states should be $2^{++}$, while the $X(6200)$ and $X(7200)$ states prefer $0^{++}$. To determine the above states more precisely, more experimental data for the channels, such as $J/\psi\psi(2S)$, $J/\psi\psi(3770)$, di-$\psi(2S)$, are required.

\end{abstract}
\maketitle
%
%
\section{Introduction\label{sec:intro}}

In 2003, the first tetraquark candidate state $\chi_{c1}(3872)$ was discovered by Belle collaboration~\cite{Choi:2003ue}, and later confirmed by BABAR~\cite{Aubert:2004ns}, CDF~\cite{Acosta:2003zx} and D0~\cite{Abazov:2004kp} groups. The discovery of this state has inaugurated a new era of studying multiquark states that provide us a unique platform to gain more insights into the low-energy quantum chromodynamics (QCD).  Since then, many charmonium- and bottomonium-like states have been observed by various experiments and, meanwhile, intrigued intensive theoretical investigations, 
see e.g. Refs~\cite{Liu:2019zoy,Brambilla:2019esw,Yang:2020atz,Guo:2017jvc,LEBED2017143,Olsen:2017bmm} for reviews on the studies of $XYZ$ particles. 

Attempts have also been made to search for states containing only four heavy quarks by, e.g., the LHCb~\cite{Aaij:2018zrb,Aaij:2020fnh} and CMS~\cite{Sirunyan:2020txn} collaborations. Recently, the LHCb collaboration declared the observation of a narrow structure around $6.9~\rm{GeV}$ and a broad structure located in the energy range $[6.2, 6.8 ]$~\rm{GeV} in the di-$J/\psi$ invariant mass event distribution, using the proton-proton collision data at centre-of-mass (c.m.) energies of $\sqrt{s}=7,\,8 $ and $13$ \rm{TeV}~\cite{Aaij:2020fnh}. The data of the di-$J/\psi$ mass spectrum also hint a possible structure in the vicinity of $7.2~\rm{GeV}$. The narrow peak can be described by employing the Breit-Wigner parametrization and an associated $X(6900)$ state is established with significance larger than $5\sigma$. The mass and width of this state are determined to be 
\bea
{\rm{M}}[X(6900)]  &=& 6886 \pm 11 \pm 11~\rm{MeV}\ ,\\
\Gamma[X(6900)]  &=& 168 \pm 33 \pm 69~\rm{MeV} \ ,
\eea  
in the scenario where the interference between the resonant contribution and the nonresonant single parton scattering is implemented. In the case without interference, the resultant determinations are  
\bea
{\rm{M}}[X(6900)] &=& 6905 \pm 11 \pm 7~\rm{MeV},  \\  
\Gamma[X(6900)]  &=& 80 \pm 19 \pm33~\rm{MeV}.
\eea
The observation of $X(6900)$ is the first experimental evidence of the fully charmed tetraquark $T_{cc\bar{c}\bar{c}}$, in another word, a new member in the $XYZ$ particle zoo is gained. 

In fact, pioneering studies on the fully charmed states can be traced back to about four decades ago~\cite{Iwasaki:1975pv, Chao:1980dv, Ader:1981db}. In Ref.~\cite{Iwasaki:1975pv}, it is pointed out that a particle at $6~\rm{GeV}$, comprising $cc\bar{c}\bar{c}$ quarks, may exist and mainly decays into two charmoniums $J/\psi\eta_c$. The existence of $cc\bar{c}\bar{c}$ states was prognosticated by using a quark-gluon model~\cite{Chao:1980dv} as well, which resides in the energy range $6$-$7~\rm{GeV}$. States composed of only heavy quarks were also predicted in the framework of potential models~\cite{Ader:1981db}. Over the years, continuous efforts have been made to explore mutiquark states, in particular, to decode their inner properties. For instance, the tetraquark states of the type $T_{cc\bar{c}\bar{c}}$ have been intensively investigated in different theoretical approaches, such as phenomenological models~\cite{Cao:2020gul,Dong:2020nwy,Guo:2020pvt,Gong:2020bmg,Zhao:2020cfi}, quark model~\cite{Wu:2016vtq,Faustov:2021hjs,Jin:2020yjn,Lu:2020cns}, potential models~\cite{Karliner:2016zzc,Deng:2020iqw}, QCD sum rules~\cite{CHEN2017247,Zhang:2020xtb, Wan:2020fsk,Yang:2020wkh,Wang:2020ols,CHEN20201994}, nonrelativistic QCD factorization~\cite{Feng:2020qee,Ma:2020kwb,Feng:2020riv}, effective theory~\cite{Zhu:2020snb,Liu:2020tqy}, etc.

Turning back to the newly observed $X(6900)$ state, plenty of theoretical works have been accumulated aiming at studying its properties, see e.g. Refs.~\cite{Wang:2020wrp,Zhao:2020nwy,Wan:2020fsk,Yang:2020wkh,Deng:2020iqw,CHEN20201994,Guo:2020pvt,Cao:2020gul,Zhu:2020snb,Liu:2020tqy,Gong:2020bmg,Ke:2021iyh,Zhu:2020xni,Albuquerque:2021erv,Karliner:2020dta,Maciula:2020wri}. For instance, the mass of $X(6900)$ obtained by QCD sum rules~\cite{Wan:2020fsk,Yang:2020wkh} agrees well with the experimental values. The $X(6900)$ state is interpreted as an excited state both in the constituent quark model~\cite{Deng:2020iqw} and in the QCD sum rule approach~\cite{CHEN20201994}. As for its inner structure, it is pointed out in Ref.~\cite{Guo:2020pvt} that the $X(6900)$ resonance is most likely to be a compact four-charm quark state with little component of molecular nature. Nevertheless, based on spectral density function sum rule and pole counting rule, Ref.~\cite{Cao:2020gul} suggests that more experimental data are needed in order to distinguish the nature of $X(6900)$. Interestingly, it is claimed in Ref.~\cite{Zhu:2020snb} that the $X(6900)$ state can be described as a Higgs-like boson, which conveys some signal beyond standard model. It is worth noting that another fully charmed state, denoted as $X(7200)$, is found in e.g. Refs.~\cite{Cao:2020gul,Wan:2020fsk,Liu:2020tqy}, which actually is also hinted by the LHCb data as mentioned above. In this work, we intend to hunt for all possible states in the di-$J/\psi$ spectrum below $7.6$~GeV, and try to determine their $J^{PC}$ quantum numbers.

To that end, we construct the amplitude for the di-$J/\psi$ production on the $pp$ collision, in which the coupled-channel effects are taken into account via final state interactions (FSI). The \{$J/\psi J/\psi$, $J/\psi\psi(2S)$, $J/\psi\psi(3700)$, $\psi(2S)\psi(2S)$\} channels are included in the $\mathcal{T}$ matrix with the aim of reproducing the di-$J/\psi$ invariant mass spectrum, reported by the LHCb collaboration~\cite{Aaij:2020fnh}, in the range from $6.2$ to $7.6$~GeV. Of course, there are other double-charmonium channels with thresholds below $7.6$~GeV, which can couple to the di-$J/\psi$ system, such as $\eta_c\eta_c$, $h_ch_c$ and $\chi_{cJ}\chi_{cJ^\prime}$. However, their contributions are either suppressed by heavy quark spin symmetry or subordinate due to the smallness of the relevant couplings estimated by meson exchanges, as pointed out by Ref.~\cite{Dong:2020nwy}, and the readers are referred to Appendix~\ref{sec:HQSS} for detailed discussions. We further perform partial wave projection of the $\mathcal{T}$ matrix within helicity formalism. Explicit expressions for the $S$-wave amplitudes with $J^{PC}=0^{++},2^{++}$ are obtained. We assume that the invariant mass spectrum is dominated by the $S$ waves, and hence those partial waves beyond $S$-wave are neglected. 

Various fits are performed to the LHCb data with the obtained production amplitudes in $S$ wave. Due to the closeness of the $X(6900)$ peak to the $J/\psi \psi(3770)$ threshold, we first take only three channels, i.e., $\{J/\psi J/\psi, J/\psi\psi(2S), J/\psi\psi(3770)\}$, into account,  and do fits with partial-wave amplitudes of $0^{++}$ and $2^{++}$, respectively. It is found that both fits can well reproduce the experimental data in the energy range below $7.2$~GeV. A di-$J/\psi$ subthreshold state, named $X(6200)$, is discovered both in the $0^{++}$ and $2^{++}$ amplitudes. However, there is no dynamically generated state responsible for the peak around $6.9~\rm{GeV}$. Thus, we make further fits, where the $\psi(2S)\psi(2S)$ channel is now switched on and the fitting range is extended up to $7.6~\rm{GeV}$ to cover more data. Following the same procedure as the three-channel fits, we find that the $0^{++}$ and $2^{++}$ amplitudes behave differently now. Furthermore, four dynamically generated states are established. We also carry out a combined fit with both $0^{++}$ and $2^{++}$ contributions, which indicates that the two partial-wave amplitudes are simultaneously sizable.

This manuscript is organized as follows. Section~\ref{sec2} serves to illustrate the basic ingredients of our theoretical framework, including coupled-channel potentials, partial-wave projection, unitarized amplitudes and production amplitude of di-$J/\psi$. Numerical results are shown in Sec.~\ref{sec3}. A brief summary is given in Sec.~\ref{sec:sum}. The explicit expressions of the helicity amplitudes are collected in Appendix ~\ref{sec:helicityamp}.

\section{Theoretical framework}\label{sec2}
In this section, the $S$-wave production amplitude, with $0^{++}$ and $2^{++}$ quantum numbers, is constructed in order to reproduce the recent di-$J/\psi$ invariant mass spectrum.

\subsection{Coupled channel potentials}
The coupled-channel effects are implemented in the production amplitude through FSI, and four channels \{$J/\psi J/\psi$, $J/\psi\psi(2S)$, $J/\psi\psi(3770)$, $\psi(2S)\psi(2S)$\} are under our consideration, as discussed in the Introduction. The relevant effective Lagrangian reads
\bea\label{eq:coneffLag}
\mathcal{L}_{\rm eff.}&=& h_1 (J/\psi\cdot J/\psi)^2\nonumber\\
&+&h_2 (J/\psi\cdot J/\psi) (J/\psi\cdot\psi(2S))\nonumber\\
&+&h_3 (J/\psi \cdot J/\psi)(J/\psi\cdot\psi(3770))\nonumber\\
&+&h_4 (J/\psi\cdot\psi(2S))^2\nonumber\\
&+&h_4^\prime (J/\psi\cdot J/\psi)(\psi(2S)\cdot\psi(2S))\nonumber\\
&+&h_5 (J/\psi\cdot\psi(2S))(J/\psi\cdot\psi(3770))\nonumber\\
&+&h_5^\prime (J/\psi\cdot J/\psi)(\psi(2S)\cdot\psi(3770))\nonumber\\
&+&h_6 (J/\psi\cdot\psi(3770))^2\nonumber\\
&+&h_6^\prime (J/\psi\cdot J/\psi)(\psi(3770)\cdot\psi(3770))  \nonumber \\ 
& +& h_7 (J/\psi \cdot \psi(2S) )(\psi(2S) \cdot \psi(2S))  \nonumber \\
&+ &h_8 (J/\psi \cdot \psi(3770) )(\psi(2S) \cdot \psi(2S)) \nonumber \\
&+& h_8^\prime (J/\psi \cdot \psi(2S) )(\psi(3770) \cdot \psi(2S))  \nonumber  \\
&+& h_9 (\psi(2S) \cdot \psi(2S))^2  ,  \
\eea
with $h_{i=1,\dots,9}$ and $h_{j=4,5,6,8}^\prime$ being the coupling constants of the 4-vector contact interactions.\footnote{In Appendix~\ref{sec:HQSS}, an effective Lagrangian in heavy quark formalism is also constructed and its connection to the above relativistic Lagrangian is discussed. }

The scattering potential for the process of $V_1(p_1,\epsilon_1)+V_2(p_2,\epsilon_2)\to V_3(p_3,\epsilon_3)+V_4(p_4,\epsilon_4)$ has the following generic form
\bea\label{eq:amp}
V_{ij} = \mathcal{C}_1\epsilon_1\cdot \epsilon_2\epsilon_3^\dagger\cdot \epsilon_4^\dagger+\mathcal{C}_2\epsilon_1\cdot \epsilon_3^\dagger\epsilon_2\cdot \epsilon_4^\dagger+\mathcal{C}_3\epsilon_1\cdot \epsilon_4^\dagger\epsilon_2\cdot \epsilon_3^\dagger\ ,\quad
\eea
where the coefficients $\mathcal{C}_k$ ($k=1,2,3$) are compiled in Table~\ref{tab:ciex}. The subscripts of $V_{ij}$ are channel labels, and $i,j\in\{1,2,3,4\}$ with the numbers in the bracket being specified by $1=[J/\psi J/\psi]$, $2=[\psi(2S) J/\psi]$ ,$3=[\psi(3770) J/\psi]$ and $4=[\psi(2S) \psi(2S)]$.

\begin{table}[htbp]
\centering
\caption{\label{tab:ciex}Coefficients of the scattering potentials.}
\begin{tabular}{r c| ccc}
\hline\hline
 $V_{ij}$& Channels & $\mathcal{C}_1$ & $\mathcal{C}_2$  & $\mathcal{C}_3$ 
 \\
\hline
11& $J/\psi J/\psi\to J/\psi J/\psi$   & $8h_1$   & $8h_1$ &$8h_1$ 
\\
12& $J/\psi J/\psi\to \psi(2S)J/\psi $   & $2h_2$    & $2h_2$  &$2h_2$ 
\\
13& $J/\psi J/\psi\to \psi(3770)J/\psi $         & $2h_3$      & $2h_3$ &$2h_3$ 
\\
 $14$ &$ J/\psi J/\psi \to \psi(2S) \psi(2S) $        & $ 4h_4^\prime$     &  $ 2h_4$                     & $ 2 h_4 $        \\
22& $ \psi(2S)J/\psi\to  \psi(2S)J/\psi$           & $2h_4$   & $4h_4^\prime$  &$2h_4$ 
\\
23& $\psi(2S)J/\psi\to  \psi(3770)J/\psi$       & $h_5$     & $2h_5^\prime$       &$h_5$ 
\\
$24$ & $ \psi(2S) J/\psi  \to  \psi(2S) \psi(2S) $        & $ 2h_7$     &  $ 2h_7$                      & $ 2h_7$       \\
 33&$\psi(3770)J/\psi\to \psi(3770)J/\psi$            & $2h_6$    & $4h_6^\prime$      &$2h_6$ 
 \\
$34$ &$ \psi(3770) J/\psi  \to  \psi(2S) \psi(2S) $        & $ 2h_8$     &  $ h_8^\prime$                      & $ h_8^\prime$       \\
$44$ & $ \psi(2S) \psi(2S) \to  \psi(2S) \psi(2S) $        & $ 8h_9$     &  $ 8h_9$                      & $ 8h_9$       \\
\hline\hline
\end{tabular}
\end{table}
\subsection{Partial wave projection}
The helicity amplitudes are defined by
\bea\label{eq:helicityamp}
V_{\lambda_1\lambda_2\lambda_3\lambda_4}= \epsilon_3^{\rho\dagger}(p_3,\lambda_3)\epsilon_4^{\sigma\dagger}(p_4,\lambda_4) V_{\mu\nu\rho\sigma} \epsilon_1^\mu(p_1,\lambda_1)\epsilon_2^\nu(p_2,\lambda_2)\ ,\nonumber\\
\eea
where $\lambda_i=\{\pm1,0\}$ are helicity eigenvalues of the polarization vectors $\epsilon_i$'s. For brevity, the channel indices are suppressed here.  There are 81 helicity amplitudes in total, however, only 25 of which are independent after imposing $P$-parity and $T$-reversal invariances. For explicit expressions of the 25 independent helicity amplitudes in our case, see Appendix~\ref{sec:helicityamp}. 

For a given angular momentum $J$, the partial wave projection of the helicity amplitudes can be obtained via
\bea
V_{\lambda_1\lambda_2\lambda_3\lambda_4}^J(s)=\frac{1}{2}\int_{-1}^{1} d z_s V_{\lambda_1\lambda_2\lambda_3\lambda_4}(s,t(s,z_s))\,d^J_{\lambda\lambda^\prime}(z_s)\,\quad
\eea
with $\lambda=\lambda_1-\lambda_2$ and $\lambda^\prime=\lambda_3-\lambda_4$. Here $s=(p_1+p_2)^2$ and $t=(p_1-p_3)^2$ are Mandelstam variables. Furthermore, $d^J_{\lambda\lambda^\prime}(z_s)$ are standard Wigner functions, $z_s\equiv\cos\theta$ and $\theta$ is the scattering angle. 
Hereafter, we focus on the S wave, therefore, $L=0$ and  $J=S$ with $S$ denoting the total spin of the initial or final states. The generalized Bose symmetry for identical particles and the conservation of $J^{PC}$ quantum numbers imply that $L+S$ should be even, which means that $J$ should take values of $0$ or $2$.

 For $S$ wave, the partial wave amplitudes in the $LJS$ basis can be obtained through
\bea
\mathcal{V}^J = \sum_{\lambda_1\lambda_2\atop\lambda_3\lambda_4}U_{\lambda_3\lambda_4}^{J}\mathcal{V}_{\lambda_1\lambda_2\lambda_3\lambda_4}^J[U_{\lambda_1\lambda_2}^{J}]^\dagger\ ,
\eea
where the transformation matrix regarding the initial states is given by
\bea
U^{J}_{\lambda_1\lambda_2}=\frac{1}{\sqrt{2S+1}}\langle S_1\lambda_1S_2-\lambda_2|S\lambda\rangle\ .
\eea
Here $S_1$ and $S_2$ are spins of the incoming vector particles $V_1$ and $V_2$, respectively.  Clesch-Gordon coefficients are represented by $\langle S_1\lambda_1S_2-\lambda_2|S\lambda\rangle$. The transformation matrix concerning the final states can be obtained exactly in the same way.

Finally, the partial wave with $J^{PC}=0^{++}$ reads
\bea
V(0^{++})=\frac{1}{3}\left(2V_{++++}^{J=0}+2V_{++--}^{J=0}+V_{0000}^{J=0}-4V_{++00}^{J=0}\right)\ . \quad
\eea
Likewise, the partial wave with $J^{PC}=2^{++}$ reads
\bea
V(2^{++})&=&\frac{2}{15}V_{0000}^{J=2}+\frac{4}{15}V_{00++}^{J=2}+\frac{4\sqrt{6}}{15}V_{00+-}^{J=2}\nonumber\\
&+&\frac{2\sqrt{6}}{15}(V_{+-++}^{J=2}+V_{-+++}^{J=2})+\frac{4\sqrt{3}}{15}(V_{00+0}^{J=2}+V_{000+}^{J=2})\nonumber\\
&+&\frac{2\sqrt{3}}{15}(V_{+0++}^{J=2}+V_{0+++}^{J=2}+V_{0-++}^{J=2}+V_{-0++}^{J=2})\nonumber\\
&+&\frac{1}{5}(V_{+0+0}^{J=2}+V_{-0+0}^{J=2}+V_{0+0+}^{J=2}+V_{0-0+}^{J=2})\nonumber\\
&+&\frac{2\sqrt{2}}{5}(V_{+-+0}^{J=2}+V_{0++-}^{J=2}+V_{-++0}^{J=2}+V_{0-+-}^{J=2})\nonumber\\
&+&\frac{2}{5}(V_{0++0}^{J=2}+V_{0-+0}^{J=2}+V_{+-+-}^{J=2}+V_{-++-}^{J=2})\nonumber\\
&+&\frac{1}{15}(V_{++++}^{J=2}+V_{--++}^{J=2}) \ .
\eea

\subsection{Unitarized amplitudes}
The conservation of probability implies a realistic amplitude should be unitary. This can be achieved by applying the on-shell-approximation version of the Bethe-Salpeter  equation~\cite{Oller:1997ti,Oller:2000fj} to the partial wave amplitudes derived in the above subsection; see Refs.~\cite{Oller:2019opk,Yao:2020bxx,Oller:2020guq} for recent reviews on various unitarization approaches. In this manner, the unitarized amplitude is given by
\bea
\mathcal{T}^J(s)=\mathcal{V}^J(s)\cdot\left[1-\mathcal{G}(s)\cdot\mathcal{V}^J(s)\right]^{-1}\ ,
\eea
with
\bea
 \mathcal{V}^J(s)&=&
\left(
\begin{array}{ccc}
 V_{11}^J(s) & \cdots &V_{1N}^J(s) \\
  \vdots & \ddots & \vdots \\
   V_{N1}^J(s) & \cdots &V_{NN}^J(s) \\
 \end{array}
\right)\ ,
\eea
where the subscripts of the entries of the matrix are channel indices. $\mathcal{G}(s)$ is a diagonal matrix defined by
\bea\label{eq:G}
\mathcal{G}(s)&=&{\rm diag}\{g_i(s)\}\ ,
\eea
 with its elements given by~\cite{Oller:1998zr}
\bea
g_i(s)&=&\frac{1}{16\pi^2}\bigg\{{a}(\mu)+\ln\frac{M_{V_1}^2}{\mu^2}
+\frac{s-\Delta}{2s}\ln\frac{M_{V_2}^2}{M_{V_1}^2}\nonumber\\
&+&\frac{\sigma(s)}{2s}\big[\ln\left(\sigma(s)+s+\Delta\right)-\ln\left(\sigma(s)-s-\Delta\right)\nonumber\\
&+&\ln\left(\sigma(s)+s-\Delta\right)-\ln\left(\sigma(s)-s+\Delta\right)\big]\bigg\}\ .\label{eq:gi}
\eea
Here $\sigma(s)=\{[s-(M_{V_1}+M_{V_2})^2][s-(M_{V_1}-M_{V_2})^2]\}^{1/2}$, $\Delta=M_{V_1}^2-M_{V_2}^2$,
and $a({\mu})$ is a subtraction constant defined at the renormalization scale of $\mu$.

Bound, virtual and resonant states correspond to pole singularities in different Riemann sheets (RSs) of the unitary $\mathcal{T}^J$ matrix.  RSs can be defined by performing analytical continuation of the $g_i(s)$ function through
\bea
g_i(s)\to g_i(s)+i\frac{p_i(s)}{4\pi \sqrt{s}}\xi_i\ ,
\eea
with $\xi_i=0,1$ and $p_i(s)$ being the c.m. momentum in the $i$th channel. Furthermore, each RS can be denoted by a number 
\bea
n=1+\sum_{i=1}^N\xi_i2^{i-1}\ ,
\eea
where $i$ is channel index and $N$ is the total number of the involved channels. In practice, the $n$th RS is often marked by Roman numerals. For example, the fourth RS is equivalent to RS-\RNum{4}.

\subsection{Production amplitude}
\begin{figure*}[htbp]
\begin{center}
\epsfig{file=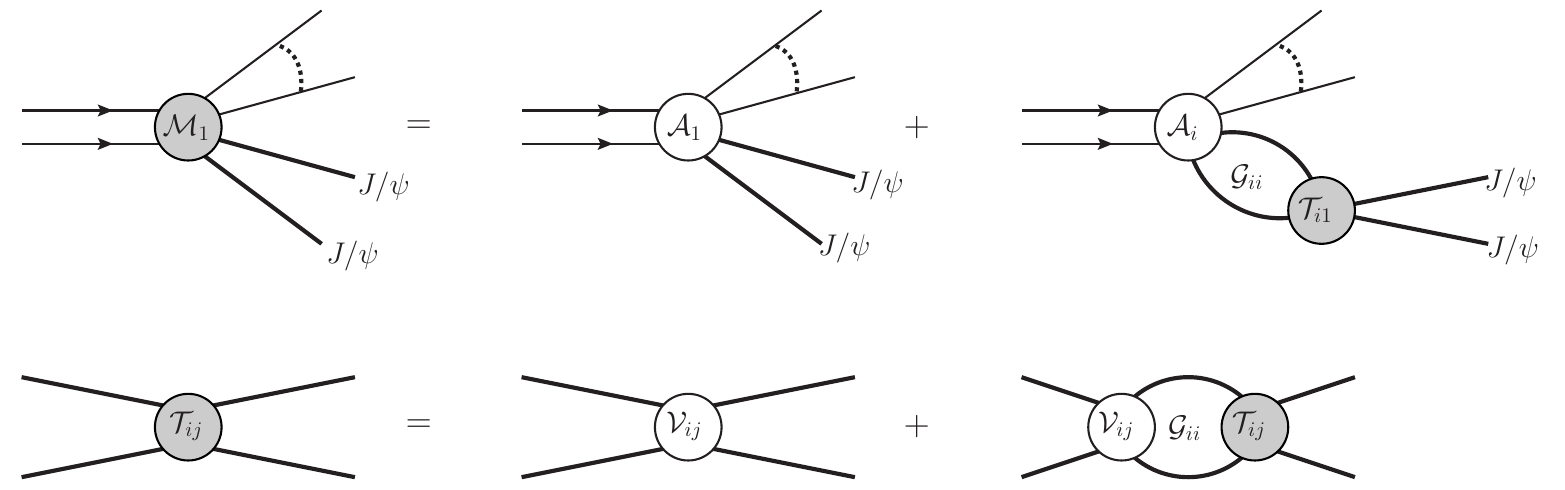,scale=0.5}
\end{center}
\caption{\label{fig:production}Diagrammatic representation of the di-$J/\psi$  production via the process of $pp\to [J/\psi J/\psi]+{anything}$ at LHCb. Coupled channel effects of $\{J/\psi J/\psi,J/\psi\psi(2S),J/\psi\psi(3700),\psi(2S)\psi(2S)\}$ are incorporated through FSI.}
\end{figure*}

The di-$J/\psi$ production amplitude, with given quantum numbers $J^{PC}$, can be written as
\bea
\mathcal{M}_1(s)&=&\mathcal{A}_1+\sum_{i} \mathcal{A}_i \mathcal{G}_{ii}(s) \mathcal{T}_{i1}(s)\ .
\eea
Here $\mathcal{G}$ is given by Eq.~\eqref{eq:G}; $\mathcal{A}_i$'s represent direct productions of the di-$J/\psi$, $J/\psi\psi(2S)$ and $J/\psi\psi(3700)$ states, respectively. The second term on the right-hand side of the above equation stands for FSI where coupled channels rescattering effects of $\{J/\psi J/\psi,J/\psi\psi(2S),J/\psi\psi(3700), \psi(2S) \psi(2S)\}$ are implemented.  A diagrammatic representation of such a production mechanism is displayed in Fig.~\ref{fig:production}. The production amplitude $\mathcal{M}_1(s)$ can be rewritten as
\bea\label{eq:proamp2}
\mathcal{M}_1(s)=\mathcal{A}_1\big[1+\sum_{i} r_i \mathcal{G}_{ii}(s) \mathcal{T}_{i1}(s)\big]\ .
\eea
The parameters $r_i$'s are defined by $r_i=\mathcal{A}_i/\mathcal{A}_1$ and $r_1=1$. In principle, $r_2$ and $r_3$ are unknown complex parameters. 

Eventually, the invariant mass of the double-$J/\psi$ spectrum is proportional to the amplitude squared multiplied by a phase space factor $\rho(s)$, namely
\bea\label{eq:forinv}
\frac{{\rm d}\mathcal{N}}{{\rm d}\sqrt{s}}\propto\rho(s)|\mathcal{M}_1(s)|^2\ ,
\eea
where the explicit expression of $\rho(s)$ is
\bea
\rho(s) = \frac{p_1(s)}{8\pi \sqrt{s}}
\ ,\quad 
p_1(s)=\frac{\lambda^{1/2}(s,m^2_{J/\psi},m^2_{J/\psi})}{2\sqrt{s}}\ .
\eea
Here $\lambda(x,y,z)=x^2+y^2+z^2-2xy-2xz-2yz$ is the so-called K\"all\'{e}n function.

\section{Numerical results and discussions}\label{sec3}
The LHCb collaboration has recently reported the invariant mass spectrum of $J/\psi$ pairs, coming from proton-proton collision data at $7,\,8$ and $13$ TeV~\cite{Aaij:2020fnh}. In general, the di-$J/\psi$ mass spectrum is dominated by the $S$-wave interaction, however, there are two $S$-wave candidates with different $J^{PC}$ quantum numbers either being $0^{++}$ or $2^{++}$. Since we have both the $0^{++}$ and $2^{++}$ amplitudes, as given in the previous section, this enables us to perform a partial wave analysis of the di-$J/\psi$ spectrum data and reveal possible underlying states with definite $J^{PC}$.  Therefore, in what follows, we are going to confront our theoretical model, given in Eq.~\eqref{eq:forinv}, with the recent LHCb data. 

\subsection{Fits with three-coupled channels}
We first consider the case with the three channels of $\{J/\psi J/\psi$, $J/\psi\psi(2S)$, $J/\psi\psi(3770)\}$, which, in principle, are sufficient to reproduce the data in the energy region below $7.2$~GeV where the structure around $6.9$~GeV is covered.  With Eqs.~\eqref{eq:proamp2} and~\eqref{eq:forinv}, the invariant mass spectrum of di-$J/\psi$ is described by
\bea\label{eq:inv2}
\frac{{\rm d}\mathcal{N}}{{\rm d}\sqrt{s}}=\rho(s)|\mathcal{A}_1(s)|^2\bigg|\gamma+\sum_{i=1}^{3} \mathcal{G}_{ii}(s) \mathcal{T}_{i1}(s)\bigg|^2\ .
\eea
It should be emphasized that the unit in Eq.~\eqref{eq:proamp2} is replaced by a constant $\gamma$ in order to simulate the coherent background. Furthermore, $r_2$ and $r_3$ are set as $1$,  such that the number of free parameters is reduced. Actually, the effects of $r_2$ and $r_3$ may be partly absorbed by the $\gamma$ constant and other free parameters involved in the rescattering amplitudes $ \mathcal{T}_{i1}(s)$.  

In Eq.~\eqref{eq:inv2}, the amplitude $|\mathcal{A}_1(s)|^2$ stands for direct production of $J/\psi$ pairs, which encodes the information on short-distance interactions. Following Ref.~\cite{Dong:2020nwy}, its modulus squared can be parametrized as
\bea\label{eq:A1}
|\mathcal{A}_1(s)|^2=\alpha^2 e^{-2\beta s}\ ,
\eea
where $\alpha$ and $\beta$ are unknown overall constant and slope parameter, respectively. The distribution of events corresponding to the double-parton scattering (DPS)~\cite{Calucci:1997ii,DelFabbro:2000ds,Calucci:1999yz}, shown in the LHCb paper~\cite{Aaij:2020fnh}, can be described by $|\mathcal{A}_1(s)|^2$ multiplied by the phase-space factor $\rho(s)$, which determines $\alpha=134$ and $\beta=0.0123$. Nevertheless, the overall constant is released as a free fitting parameter in our fits to be discussed below.

Coupled channel effects are implemented through FSI by the summation term in Eq.~\eqref{eq:inv2}. We employ a uniform subtraction constant $a(\mu)$ for all the two-point $\mathcal{G}_{ii}(s)$ functions.\footnote{In principle, one should adopt different values for the subtraction constants $a(\mu)$ in different channels. However, this could complicate the production amplitudes; see Ref.~\cite{Oller:2002na} for more discussions.} Besides, the subtraction constant is fixed, i.e., $a(\mu=1~{\rm GeV}) = -3.0$, where the renormalization scale is set to $1$~GeV. We have checked that $a(\mu)$ tends to take the value of $-3$ even if it is set as a free fitting parameter. 
In fact, this value also corresponds closely to the natural value of a subtraction constant~\cite{Oller:2000fj,Guo:2020pvt},  which can be  directly calculated by using the formula derived in Ref.~\cite{Guo:2018tjx}. Note that we are using the same value for the subtraction constant for all the fits performed in this work. 
As for the scattering amplitudes $\mathcal{T}_{i1}(s)$, there are a few coupling constants stemming from the effective Lagrangian. To reduce the number of fitting parameters and to obtain natural values for those couplings, we employ
\bea
h_{i}=h_{i}^\prime\ ,\qquad {h}_i=\bar{h}_i\cdot{\sum_{j=1}^4\sqrt{2m_{j}}}\ ,
\eea
where $m_j$ are masses of the vector particles involved in the scatterings, as specified in Eq.~\eqref{eq:processes}. The values of the masses are taken from PDG~\cite{Zyla:2020zbs}. In this way, the $\bar{h}_i$'s are chosen to be fitting parameters instead of ${h}_i$'s. Finally, there are in total eight free parameters: $\bar{h}_1,\dots,\bar{h}_6$, $\alpha$ and $\gamma$. For the coupled-channel case of $\{J/\psi J/\psi$, $J/\psi\psi(2S)$, $J/\psi\psi(3770)\}$, two different kinds of fits are performed, which are denoted by Fit-A and Fit-B.

\begin{table}[tbp]
\caption{\label{tab:fit}Results of fits with three-coupled channels. The asterisk denotes an input quantity.}
\centering
\begin{tabular}{c|rr}
\hline
 & Fit-A ($0^{++}$) & Fit-B ($2^{++}$)\\
\hline
 $\bar{h}_1$ & $0.13^{+0.06}_{-0.02}$ & $0.34^{+0.16}_{-0.06}$\\
 $\bar{h}_2$ & $1.0^{+0.4}_{-0.2}$ & $2.6_{-0.5}^{+1.0}$\\
 $\bar{h}_3$ & $0.01^{+0.11}_{-0.11}$  & $-0.02_{-0.27}^{+0.28}$\\ 
 $\bar{h}_4$ & $1.3^{+0.4}_{-0.3}$  &$2.5_{-0.5}^{+0.9}$\\
 $\bar{h}_5$ & $-0.43^{+0.29}_{-0.25}$ & $-0.85_{-0.5}^{+0.6}$\\
 $\bar{h}_6$ & $-0.19^{+0.09}_{-0.05}$ &$-0.38_{-0.11}^{+0.18}$ \\
 $\alpha$ & $219^{+82}_{-18}$ & $222^{+84}_{-21}$\\
 $\beta$ &$0.0123^*$&$0.0123^*$\\
 $\gamma$ & $-0.23^{+0.30}_{-0.08}$ & $-0.22_{-0.09}^{+0.30}$\\
\hline
$\chi^2/{\rm d.o.f.}$ & $\frac{26.8}{36-8}\simeq 0.96$ & $\frac{26.9}{36-8}\simeq 0.96$\\
\hline
\end{tabular}
\end{table}

\begin{figure}[tbp]
\begin{center}
\epsfig{file=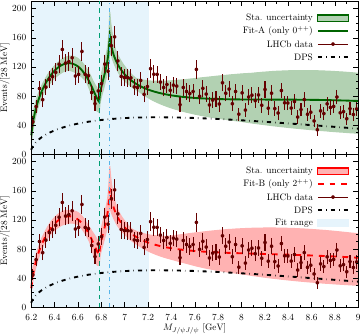,scale=1.4}
\end{center}
\caption{\label{fig:fit.3cc} Description of the LHCb data of the di-$J/\psi$ invariant mass distribution with three-coupled channels $\{J/\psi J/\psi$, $J/\psi\psi(2S)$, $J/\psi\psi(3770)\}$. The fit range is from $6.2$ to $7.2$~GeV, indicated by the light blue range. The error bands are obtained by varying the coupling constants within their 1-$\sigma$ uncertainties.}
\end{figure}

In Fit-A, the di-$J/\psi$ production amplitude is completely saturated by the contribution from the partial wave with $J^{PC}=0^{++}$.  Results of Fit-A are shown in the second column of Table~\ref{tab:fit} and in the upper panel of Fig.~\ref{fig:fit.3cc}. Fit-A is performed up to $7.2$~GeV. Excellent agreement between our model and the LHCb data is achieved with a $\chi^2/{\rm d.o.f.}\simeq 0.96$. The obtained values of the coupling constants $\bar{h}_i$ turn out to be of natural size as expected. It can also be seen from the figure that, the di-$J/\psi$ spectrum is well described up to $9$~GeV, even though the fitting range of Fit-A is from $6.2$ to $7.2$~GeV.

Fit-B corresponds to the case that the production amplitude consists of the partial wave with $J^{PC}=2^{++}$ only. The fitting range of Fit-B is the same as that of Fit-A.  Results of Fit-B are compiled in the third column of Table~\ref{tab:fit}, and plots are displayed in the lower panel of Fig.~\ref{fig:fit.3cc}. It can be found that the fit quality of Fit-B is as good as Fit-A below $7.2$~GeV, and the data beyond the fitting range are again well described within 1-$\sigma$ uncertainty.

Since the coupling constants involved in the potentials are determined by the fits, it is now ready for us to investigate the pole structures of the unitarized amplitude in the RS. The pole positions and their residues are obtained using the fitted values of parameters of Fit-A and Fit-B, which are collected in Tables~\ref{tab:poleA} and~\ref{tab:poleB}, respectively. In each case, a near-threshold bound state is found:
$\sqrt{s}_{\rm pole}=6173.9_{-41.5}^{+19.6}~\rm{MeV}$ 
for the fit with partial wave of $0^{++}$ and 
$\sqrt{s}_{\rm pole}=6169.3^{+23.9}_{-44.2}~\rm{MeV}$ 
for the fit with partial wave of $2^{++}$. This state is referred to as $X(6200)$ in Ref.~\cite{Dong:2020nwy}. Compared with the three-coupled potentials used in Ref.~\cite{Dong:2020nwy}, we systematically incorporate energy dependent terms via an effective Lagrangian approach. We conclude that the inclusion of those terms does not affect the existence of the $X(6200)$ bound state. However, the LHCb data of the di-$J/\psi$ invariant mass distribution in the range $[6.2,\,7.2~{\rm GeV}]$ are described almost equally well by the $0^{++}$ and $2^{++}$ fits. Namely, the difference between the $0^{++}$ and $2^{++}$ partial wave amplitudes is not sensitive to the events distributed in the fitting range at all, preventing us from disentangling the $J^{PC}$ quantum numbers of this state.

The peak around $6.9$~GeV is well described both in Fit-A and Fit-B, however, to which the dominate contribution is from the effect of the $J/\psi\psi(3770)$ threshold. No resonant poles, responsible for this peak, were found in the nearby energy region. In Tables~\ref{tab:poleA} and~\ref{tab:poleB}, the  RS-\RNum{2} poles are presented for easy comparison with the results in Ref.~\cite{Dong:2020nwy}.

\begin{table}[tbp]
\caption{\label{tab:poleA}Poles and their residues based on Fit-A. The RSs, on which the poles are located, are given in the first column.  }
\centering
\begin{tabular}{c|c|ccc}
\hline
 & Position &\multicolumn{3}{c}{$|{\rm Residue}|^{1/2}$~[GeV]}\\
RS&$\sqrt{s_{\rm pole}}$~[MeV]&$J/\psi J/\psi$&$J/\psi\psi(2S)$&$J/\psi\psi(3770)$\\
\hline
$\rm{\RNum{1}}$ & $6173.9_{-41.5}^{+19.6}$ & $16.8_{-10.3}^{+5.8}$& $22.6_{-11.3}^{+4.1}$& $5.1_{-2.8}^{+3.7}$\\
$\rm{\RNum{2}}$ & $6191.4_{-4.8}^{+2.4}$ & $4.2_{-1.0}^{+1.3}$ &  $5.7_{-1.0}^{+1.2}$ &  $1.3_{-0.7}^{+0.9}$\\
$\rm{\RNum{2}}$  & $6976.2^{+70.6}_{-75.5}-i153.0_{-123.3}^{+182.6}$& $27.6^{+8.1}_{-7.1}$ & $36.2^{+6.8}_{-6.7}$ & $19.5^{+10.9}_{-6.3}$ \\
\hline
\end{tabular}
\end{table}

\begin{table}[tbp]
\caption{\label{tab:poleB} Poles and their residues based on Fit-B. The RSs, on which the poles are located, are given in the first column. }
\centering
\begin{tabular}{c|c|ccc}
\hline
 & Position &\multicolumn{3}{c}{$|{\rm Residue}|^{1/2}$~[GeV]}\\
RS&$\sqrt{s_{\rm pole}}$~[MeV]&$J/\psi J/\psi$&$J/\psi\psi(2S)$&$J/\psi\psi(3770)$\\
\hline
$\rm{\RNum{1}}$  & $6169.3^{+23.9}_{-44.2}$ & $17.8^{+5.5}_{-10.4}$& $22.8^{+3.6}_{-10.3}$& $5.3^{+3.1}_{-3.0}$\\
$\rm{\RNum{2}}$ & $6190.9^{+2.8}_{-5.1}$ & $4.4^{+1.2}_{-1.2}$ &  $5.7^{+1.1}_{-1.0}$ &  $1.3^{+0.8}_{-0.7}$\\
$\rm{\RNum{2}}$ & $6991.7^{+121.6}_{-87.8}-i176.1_{-171.1}^{+291.6}$& $29.3^{+9.8}_{-8.8}$ & $38.5^{+8.7}_{-9.6}$ & $20.4^{+17.9}_{-9.2}$ \\
\hline
\end{tabular}
\end{table}

\subsection{Fits with four-coupled channels}

As discussed in the previous subsection, the LHCb data below $7.2$~GeV cannot distinguish between the partial waves of $J^{PC}=0^{++}$ and $2^{++}$ in the three-coupled cases. To tackle thus issue, one may extend the fitting range to include more data at higher energies, such that the partial wave amplitude involved in the invariant mass spectrum formula~Eq.~\eqref{eq:inv2} will be much more constrained. On the other hand, there exists a bump of events distribution just above $7.2$~GeV, as can be seen in Figure~\ref{fig:fit.3cc}, and the prediction, based on the coupled $J/\psi J/\psi$-$J/\psi\psi(2S)$-$J/\psi\psi(3770)$ model, actually fails to describe it. Thus, we improve our formulation by incorporating one more channel, the $\psi(2S)\psi(2S)$ channel, at the price of introducing three extra unknown parameters $\bar{h}_7$, $\bar{h}_8$ and $\bar{h}_9$.

\begin{table}[htbp]
\caption{\label{tab:fit4cc}Results of fits with four-coupled channels. The asterisk denotes an input quantity.}
\centering
\begin{tabular}{c|rr}
\hline
 & Fit-C ($0^{++}$) & Fit-D ($2^{++}$)\\
\hline
 $\bar{h}_1$ & $-4.7_{-1.5}^{+0.9}$  & $-6.5_{-0.6}^{+1.1}$\\
 $\bar{h}_2$ & $-16.6_{-5.4}^{+3.3}$ & $-27.7_{-1.3}^{+3.0}$\\
 $\bar{h}_3$ & $1.0_{-0.5}^{+1.0}$   & $-87.8_{-6.2}^{+13.3}$\\ 
 $\bar{h}_4$ & $-12.2_{-4.1}^{+2.5}$ &$-24.1_{-1.0}^{+1.3}$\\
 $\bar{h}_5$ & $1.5_{-0.8}^{+1.4}$   &$-99.1_{-25.1}^{+35.4}$\\
 $\bar{h}_6$ & $-0.28_{-0.15}^{+0.11}$ &$-318.3_{-115.5}^{+54.5}$ \\
 $\bar{h}_7$ & $-17.7_{-5.7}^{+3.6}$ &$-26.0_{-2.3}^{+1.7}$ \\
 $\bar{h}_8$ & $1.4_{-0.7}^{+1.3}$   &$-182.1_{-16.8}^{+9.5}$ \\ 
 $\bar{h}_9$ & $-4.7_{-1.7}^{+1.1}$ &$-5.4_{-0.8}^{+0.3}$ \\
 $\alpha$ & $1641^{+487}_{-614}$ & $186_{-41}^{+70}$\\
 $\beta$ &$0.0123^*$&$0.0123^*$\\
 $\gamma$ & $1.14^{+0.09}_{-0.04}$ & $-0.01_{-0.3}^{+0.3}$\\
\hline
$\chi^2/{\rm d.o.f.}$ & $\frac{29.3}{50-11}\simeq 0.75$ & $\frac{38.5}{50-11}\simeq 0.99$\\
\hline
\end{tabular}
\end{table}

\begin{figure}[tbp]
\begin{center}
\epsfig{file=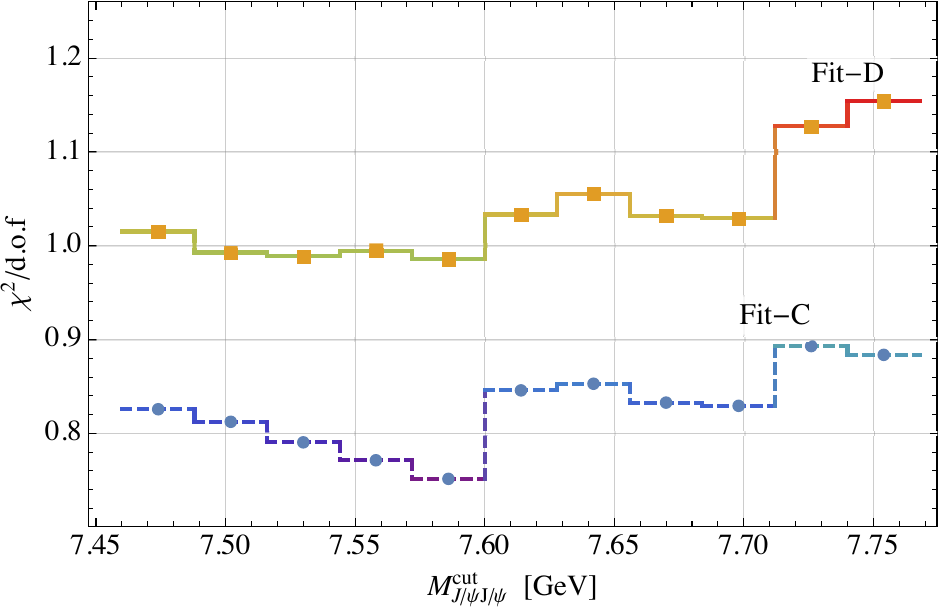,scale=0.54}
\end{center}
\caption{\label{fig:chisqs}Values of $\chi^2/{\rm d.o.f.}$ for Fit-C and Fit-D up to various maximum energies, $M_{J/\psi J/\psi}^{\rm cut}$.}
\end{figure}

\begin{figure}[tbp]
\begin{center}
\epsfig{file=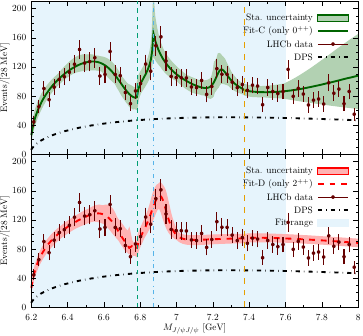,scale=1.4}
\end{center}
\caption{\label{fig:fit-4cc}Description of the LHCb data of the di-$J/\psi$ invariant mass distribution with four-coupled channels $\{J/\psi J/\psi$, $J/\psi\psi(2S)$, $J/\psi\psi(3770)$, $\psi(2S)\psi(2S)\}$. The fit range is from $6.2$ to $7.6$~GeV, indicated by the light blue range. The error bands are obtained by varying the coupling constants within their 1-$\sigma$ uncertainties.}
\end{figure}

\begin{figure}[tbp]
\begin{center}
\epsfig{file=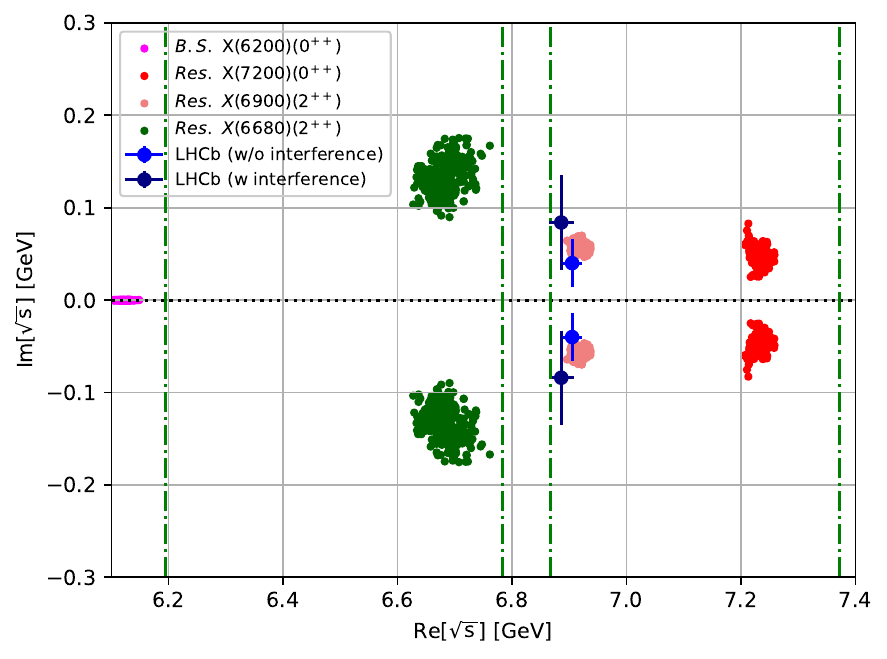,scale=0.58}
\end{center}
\caption{\label{fig:pole-4c} Poles from the four-channel fits. The green dot-dashed lines denote the positions of the thresholds.}
\end{figure}

Two new fits, denoted by Fit-C (only $0^{++}$) and Fit-D (only $2^{++}$) , are done in the procedures similar to Fit-A and Fit-B, respectively. Results of the fits are collected in Table~\ref{tab:fit4cc}. 
The fits are performed using the data of events in the energy range from the di-$J/\psi$ threshold to the energy point $7.6$~GeV, i.e. $[2m_{J/\psi},M_{J/\psi J/\psi}^{\rm cut}=7.6~{\rm GeV}]$.  The above value for $M_{J/\psi J/\psi}^{\rm cut}$ is chosen such that a plateau-like behavior for Fit-D and a downtrend for Fit-C end simultaneously, as can be seen from Fig.~\ref{fig:chisqs}. 
Compared to the three-coupled-channel fits, the fitting quality of Fit-C is improved in view of the obtained $\chi^2/{\rm d.o.f.}$ value, due to the inclusion of the fourth channel with more free coupling constants. However, we find that in the current cases the fitting parameters are now much more correlated, and consequently the resultant statistical errors become larger than those of the above three-coupled-channel fits. A comparison between the LHCb data and our predictions is shown in Fig.~\ref{fig:fit-4cc}. It can be seen from the figure that the theoretical results of invariant mass spectrum, contributed either by the $0^{++}$  partial wave or by the $2^{++}$ one,  behave quite differently now, as expected.

\begin{table*}[htbp]
\caption{\label{tab:pole4cc} Poles and their residues based on Fit-C\&D. The RSs, on which the poles are located, are given in the first column. The $J^{PC}$ quantum numbers are shown in the brackets, and here in the Table we use $0^{++}$ and $2^{++}$ to denote Fit-C and Fit-D, respectively. }
\centering
\begin{tabular}{c|c|cccc}
\hline
 & Position &\multicolumn{4}{c}{$|{\rm Residue}|^{1/2}$~[GeV]}\\
RS&$\sqrt{s_{\rm pole}}$~[MeV]&$J/\psi J/\psi$&$J/\psi\psi(2S)$&$J/\psi\psi(3770)$& $\psi(2S)\psi(2S)$\\
\hline
$\rm{\RNum{1}}$ ($0^{++}$)& $6124.8^{+23.9}_{-121.8}$& $24.6^{+7.5}_{-2.3}$ & $21.0^{+21.2}_{-6.7}$ & $1.1^{+1.1}_{-1.1}$ & $2.7^{+18.2}_{-2.6}$ \\
 $\rm{\RNum{8}}$ ($0^{++}$)& $7234.3^{+24.2}_{-28.5}-i45.1_{-20.0}^{+37.8}$& $5.6^{+2.1}_{-1.8}$ & $6.2^{+1.1}_{-1.1}$ & $0.9^{+0.3}_{-0.3}$ & $37.0^{+2.6}_{-2.3}$ \\
$\rm{\RNum{2}}~$($2^{++}$)& $6680.3^{+80.9}_{-53.0}-i136.1_{-46.4}^{+39.3}$& $14.9^{+2.1}_{-2.4}$ & $26.5^{+1.8}_{-2.8}$ & ${7.8^{+5.9}_{-1.7}}$  & $39.0^{+4.9}_{-5.2}$\\
$\rm{\RNum{8}}$ ($2^{++}$)& $6919.8^{+17.2}_{-23.7}-i58.8_{-12.2}^{+10.9}$& $5.7^{+1.6}_{-1.3}$ & $9.9^{+1.1}_{-1.8}$ & $2.9^{+1.2}_{-0.5}$  & $52.7^{+1.7}_{-1.1}$\\
\hline
\end{tabular}
\end{table*}

For Fit-C where the $0^{++}$ partial wave is used, the agreement is excellent in the sense that the theoretical line shape coincides with all the details of the LHCb data in the entire fitting range. Based on Fit-C, a pole is found in the RS-\RNum{8} and its location reads 
\bea
\sqrt{s_{\rm pole}}=(7234.3^{+24.2}_{-28.5}-i45.1_{-20.0}^{+37.8})~{\rm MeV}\ .
\eea
This resonant state is called $X(7200)$ in literature, e.g., Ref.~\cite{Cao:2020gul}. The coupling strengths of the $X(7200)$ state with the four channels, i.e., the square roots of the magnitudes of the residues, are specified in Table~\ref{tab:pole4cc}. As can be seen from the Table, this state is most strongly coupled to $\psi(2S)\psi(2S)$. The emergence of the $X(7200)$ state accounts for the bump around $7.3$~GeV in the events distribution.  In addition to the discovery of the $X(7200)$ state, a pole of 
$\sqrt{s_{\rm pole}}=6124.8^{+23.9}_{-121.8}~\rm{MeV}
$
is obtained as well. It can be regarded as the $X(6200)$ state discussed in the preceding three-coupled-channel analysis, but the pole location is shifted by an amount of about $50$~MeV due to the incorporation of the di-$\psi(2S)$ channel. It is worth noting that the above two states observed in the $0^{++}$ partial wave amplitudes are absent in the $2^{++}$ ones, providing us evidence that their $J^{PC}$ quantum numbers should be $0^{++}$. Detailed pole information on the two states is given in Table~\ref{tab:pole4cc}.

For Fit-D in which the $2^{++}$ partial wave is implemented, our prediction is in good agreement below $7.2$~GeV,  however, deviates from the events data of the bump around $7.3$~GeV. Nevertheless, the procedure of pole hunting reveals a pole structure located at 
\bea
\sqrt{s_{\rm pole}}=(6919.8^{+17.2}_{-23.7}-i58.8_{-12.2}^{+10.9})~{\rm MeV}\ .
\eea
which can be associated with the $X(6900)$ state reported by the LHCb collaboration~\cite{Aaij:2020fnh}. This finding indicates that the $J^{PC}$ quantum numbers are much more likely to be $2^{++}$. It can also be found from Table~\ref{tab:pole4cc} that the coupling strength of $X(6900)$ to the di-$\psi(2S)$ pair is much larger than the states in the other three channels. Hence, the di-$\psi(2S)$ channel is of particular importance to dynamically generate the $X(6900)$ resonant state. 
In other words, the $X(6900)$ state cannot be dynamically generated unless the heavier channel $\psi(2S) \psi(2S)$ is included. In the absence of the $\psi(2S) \psi(2S)$ channel, an alternative way of obtaining the $X(6900)$ state is to introduce a Castillejo-Dalitz-Dyson (CDD) pole in the amplitude, as done in Ref.~\cite{Guo:2020pvt}. 
It seems compatible between Ref.~\cite{Guo:2020pvt} and this study that the former requires a CDD pole and the latter finds that the resonance is mostly due to the dynamical contributions from the $\psi(2S)\psi(2S)$ channel, which is not included in Ref.~\cite{Guo:2020pvt} because of its rather large threshold compared to the $X(6900)$ mass. Namely, a CDD pole could mimic dynamically generated states by heavier channels, which are not included explicitly in the coupled-channel study.

 Interestingly, a RS-\RNum{2} pole with a large imaginary part is also achieved in the $2^{++}$ partial wave amplitude, which reads 
\bea\sqrt{s_{\rm pole}}=6680.3^{+80.9}_{-53.0}-i136.1_{-46.4}^{+39.3}~{\rm MeV}\ .
\eea
The broad structure, ranging from $6.2$ to $6.8$~GeV in the spectrum and dubbed threshold enhancement by the LHCb collaboration~\cite{Aaij:2020fnh},  probably contains also an inherent ingredient, here represented by the obtained dynamically generated pole of $J^{PC}=2^{++}$, in addition to the kinematical threshold effects. For more information on the poles we obtained, see Table~\ref{tab:pole4cc}. 

In Fig.~\ref{fig:pole-4c}, we plot the pole locations in the complex $\sqrt{s}$ plane. For comparison, the results of $X(6900)$ reported by the LHCb collaboration are also shown. Our determination of the $X(6900)$ state is consistent with the experimental values within 1-$\sigma$ uncertainties. 

\subsection{Combined fit}
It is believed that the invariant mass spectrum should be dominated by $S$-wave contributions. However, there is no {\it a priori} criterion for us to judge if the $0^{++}$ partial wave or the $2^{++}$ one is more important than the other, since both of them have the same orbital angular momentum $L=0$. It might be an appropriate way to treat them as equally important ingredients and, meanwhile, let the events data make the judgement. To that end,  the invariant mass spectrum is recast as
\bea\label{eq:inv3}
\frac{{\rm d}\mathcal{N}}{{\rm d}\sqrt{s}}=\sum_{J}\rho(s)|\mathcal{A}_1^J(s)|^2&\bigg|&\gamma_J+\sum_{i} \mathcal{G}_{ii}(s) \mathcal{T}_{i1}^{J}(s)\bigg|^2\ ,
\eea
where we have used $J=0,2$ to denote the $0^{++}$ and $2^{++}$ amplitudes in $S$ wave. Here, the incoherent backgrounds are assumed to be uniform. That is $\mathcal{A}_1^{J=0}(s)=\mathcal{A}_1^{J=2}(s)$, which are parametrized in the same form as Eq.~\eqref{eq:A1}. On the other hand, we adopt different coherent backgrounds for $J=0$ and $J=2$ cases, represented by $\gamma_0$ and $\gamma_2$ in Eq.~\eqref{eq:inv3}, respectively.

\begin{table}[tbp]
\caption{\label{tab:fit3c2}Results of the combined fit. The asterisk denotes an input quantity.}
\centering
\begin{tabular}{c|r}
\hline
 & Fit-E ($0^{++} \text{~and~} 2^{++}$)\\
\hline
 $\bar{h}_1$ & $-3.7_{-2.9}^{+0.5}$ \\
 $\bar{h}_2$ & $-15.8_{-13.3}^{+2.3}$ \\
 $\bar{h}_4$ & $-12.8_{-9.0}^{+1.7}$  \\ 
 $\bar{h}_7$ & $-14.6_{-9.4}^{+2.1}$  \\
 $\bar{h}_9$ & $-2.5_{-1.6}^{+0.4}$ \\
 $\alpha$ & $837_{-242}^{+484}$ \\
 $\beta$ &$0.0123^*$\\
 $\gamma_0$ & $1.2_{-0.1}^{+0.1}$ \\
  $\gamma_2$ & $0.9_{-0.2}^{+0.1}$ \\
\hline
$\chi^2/{\rm d.o.f}$ & $\frac{44.2}{50-8}\simeq 1.05$ \\
\hline
\end{tabular}
\end{table}

\begin{figure}[tbp]
\begin{center}
\epsfig{file=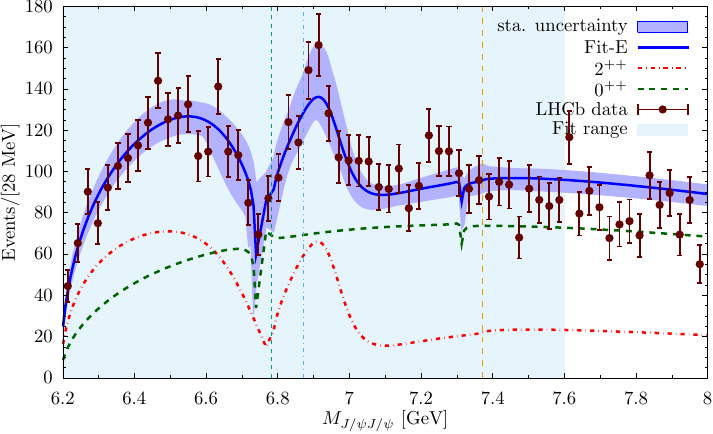,scale=0.7}
\end{center}
\caption{\label{fig:fit.com}Description of the LHCb data of the di-$J/\psi$ invariant mass distribution with three-coupled channels $\{J/\psi J/\psi$, $J/\psi\psi(2S)$, $\psi(2S)\psi(2S)\}$. The fit range is from $6.2$ to $7.6$~GeV, indicated by the light blue range. The error band is obtained by varying the coupling constants within their 1-$\sigma$ uncertainties.}
\end{figure}

\begin{table*}[htbp]
\caption{\label{tab:polecom} Poles and their residues based on the combined fit. The RSs, on which the poles are located, are given in the first column. The $J^{PC}$ quantum numbers of the obtained states are specified in the brackets.}
\centering
\begin{tabular}{r|c|ccc}
\hline
 & Position &\multicolumn{3}{c}{$|{\rm Residue}|^{1/2}$~[GeV]}\\
RS ($J^{PC}$)&$\sqrt{s_{\rm pole}}$~[MeV]&$J/\psi J/\psi$&$J/\psi\psi(2S)$&$\psi(2S)\psi(2S)$\\
\hline
$\rm{\RNum{1}}~$($0^{++}$) & $5979.6^{+3.6}_{-9.1}$ & $36.4^{+0.6}_{-0.2}$& $6.1^{+0.6}_{-1.7}$& $2.9^{+0.3}_{-0.8}$\\
$\rm{\RNum{2}} $ ($2^{++}$) &$6769.7^{+163.1}_{-140.2}-i204.5_{-62.3}^{+46.5}$ & $17.5^{+0.7}_{-1.5}$ &  $22.4^{+6.7}_{-6.9}$ &  $44.8^{+4.1}_{-4.7}$\\

$\rm{\RNum{4}}$ ($2^{++}$) & $6951.1^{+36.1}_{-48.2}-i89.1_{-17.8}^{+15.4}$& $7.2^{+1.7}_{-1.5}$ & $11.2^{+1.8}_{-2.0}$ & $50.8^{+3.0}_{-1.8}$ \\
\hline
\end{tabular}
\end{table*}

With the above preparations, fits incorporating both the $0^{++}$ and $2^{++}$ waves can be performed. It is found that, when we make a combined fit with four-coupled channels up to $7.6$~GeV,  the data around $6.9$~GeV seem not sufficient enough to distinguish the nearby threshold effect of the $J/\psi\psi(3770)$ channel from the contribution of a $X(6900)$ state. As a result, a stable solution cannot be obtained in the four-channel fit. Therefore, we exclude the threshold effect by switching off the $J/\psi\psi(3770)$ channel, and redo a combined fit (Fit-E) with the channels of $\{J/\psi J/\psi$, $J/\psi\psi(2S)$, $\psi(2S)\psi(2S)\}$. Results of the combined fit are shown in Table~\ref{tab:fit3c2} and in Fig.~\ref{fig:fit.com}. It can seen from Fig.~\ref{fig:fit.com} that both the $0^{++}$ and $2^{++}$ waves contribute sizably to the invariant mass spectrum.   The former yields a bound state, located at $\sqrt{s_{\rm pole}}=5979.6^{+3.6}_{-9.1}~\rm{MeV}$. It should correspond to the $X(6200)$ state, if considering the influence caused by the closure of the $J/\psi\psi(3770)$ channel. The latter allows the existence of a broad resonance and a narrow pole for the $X(6900)$ state, as shown in Table~\ref{tab:polecom}.

However, unlike the results of poles in Fit-C, there is no pole structure that accounts for the event enhancement around $7.3$~GeV now.  We owe the absence of such pole to the fact that the  $J/\psi\psi(3770)$ channel is excluded, which could have considerable impact on the formation of this pole. More data from experiments are indispensable for performing a comprehensive combined fit with four channels of  $\{J/\psi J/\psi$, $J/\psi\psi(2S)$, $J/\psi\psi(3770)$, $\psi(2S)\psi(2S)\}$, in order to draw a solid conclusion on the existence of the $X(7200)$ state discovered in Fit-C.

 \section{Summary and conclusions\label{sec:sum}}

In this work, we have presented a partial wave analysis of the recent di-$J/\psi$ invariant mass spectrum. Coupled-channel effects, due to the rescatterings among the $\{J/\psi J/\psi$, $J/\psi\psi(2S)$, $J/\psi\psi(3770)$, $\psi(2S)\psi(2S)\}$ states, are included in the production amplitude of $pp\to J/\psi J/\psi+anything$. The unitary $S$-wave amplitudes with $J^{PC}=0^{++}$ and $2^{++}$ are obtained with the help of the Bethe-Slapeter equation under on-shell approximation. Various fits are performed and the di-$J/\psi$ invariant mass spectrum can be well reproduced. Nevertheless, it is found that, when fits are performed up to $7.6$~GeV, the $0^{++}$ and $2^{++}$ amplitudes behave differently, enabling one to determine the $J^{PC}$ quantum numbers of the dynamically generated poles. Our final results are based on the four-coupled channel fits. Four states are found. In the case of $0^{++}$ wave, a bound state $X(6200)$ is located below the di-$J/\psi$ threshold and a narrow resonant state $X(7200)$ resides at 
$
\sqrt{s_{\rm pole}}=(7234.3^{+24.2}_{-28.5}-i45.1_{-20.0}^{+37.8})~{\rm MeV}.
$
In the case of $2^{++}$ wave, a broad resonant state $X(6680)$ is discovered with 
$
\sqrt{s_{\rm pole}}=(6680.3^{+80.9}_{-53.0}-i136.1_{-46.4}^{+39.3})~{\rm MeV}
$
and a narrow resonant state $X(6900)$ exists with
$
\sqrt{s_{\rm pole}}=(6919.8^{+17.2}_{-23.7}-i58.8_{-12.2}^{+10.9})~{\rm MeV}.
$
These findings can be determined more precisely in the future when more experimental data in the channels of $J/\psi\psi(2S)$, $J/\psi\psi(3770)$ and $\psi(2S)\psi(2S)$ are available.

 \acknowledgments
We would like to thank F.-K.~Guo, Z.-H.~Guo and S.-L.~Zhang for helpful discussions. We are also grateful to the referee for inspiring us to make a comparison between the inclusion of the $\psi(2S)\psi(2S)$ channel and the introduction of a CDD pole. This work is supported by National Nature Science Foundations of China (NSFC) under Contract No. 11905258 and by the Fundamental Research Funds for the Central Universities.

\appendix
\section{Helicity amplitudes\label{sec:helicityamp}}
In our case, it is straightforward to obtain that $V_{\mu\nu\rho\sigma}=\mathcal{C}_1g_{\mu\nu}g_{\rho\sigma}+\mathcal{C}_2g_{\mu\rho}g_{\nu\sigma}+\mathcal{C}_3g_{\mu\sigma}g_{\nu\rho}$ by comparing Eq.~\eqref{eq:helicityamp} with Eq.~\eqref{eq:amp}. Thus, for the process of 
\bea\label{eq:processes}
\hspace{-0.7cm}V_1(p_1,m_1)+V_2(p_2,m_2)\to V_3(p_3,m_3)+V_4(p_4,m_4),
\eea 
one can derive explicitly the 25 independent helicity amplitudes, which are listed below.

\bea
V_{0000}&=&\frac{1}{m_1 m_2 m_3 m_4} \bigg[ \mathcal{C}_3 (p_{cm} \bar{p}_{cm} +z_s \omega_2 \omega_3)(p_{cm} \bar{p}_{cm}   \nonumber \\
&&+ z_s \omega_1 \omega_4)+ \mathcal{C}_2(p_{cm} \bar{p}_{cm} - z_s \omega_1 \omega_3) (p_{cm} \bar{p}_{cm}   \nonumber \\
&&-  z_s \omega_2 \omega_4)+\mathcal{C}_1 (p_{cm}^2+\omega_1 \omega_2) (\bar{p}_{cm}^2 + \omega_3 \omega_4) \bigg] \ ,  \nonumber \\
V_{00++}
 &=& \frac{1}{2m_1 m_2}\bigg[ (\mathcal{C}_2 + \mathcal{C}_3)(-1+z_s^2) \omega_1 \omega_2 \nonumber\\
 &&\hspace{1.5cm}- 2 \mathcal{C}_1 (p_{cm}^2 + \omega_1 \omega_2) \bigg]  \ ,\nonumber \\
V_{00+-}
&=& -\frac{1}{2m_1 m_2} \bigg[ (\mathcal{C}_2 + \mathcal{C}_3) (-1+z_s^2) \omega_1 \omega_2 \bigg]  \ , \nonumber \\
V_{++++}
&=&\mathcal{C}_1+\frac14 \bigg[ \mathcal{C}_3(-1+z_s)^2+\mathcal{C}_2(1+z_s)^2 \bigg] \ ,  \nonumber  \\
V_{--++}
&=&  \mathcal{C}_1 + \frac14 \bigg[ \mathcal{C}_2 (-1+z_s)^2 + \mathcal{C}_3 (1+z_s)^2 \bigg]  \ , \nonumber \\
V_{+0++} 
& =&  -\frac{\sqrt{1-z_s^2}}{2\sqrt{2} m_2} \bigg[ \mathcal{C}_3(-1+z_s)+\mathcal{C}_2 (1+z_s) \bigg] \omega_2
 \ ,  \nonumber  \\ 
V_{0+++}
&=& \frac{\sqrt{1-z_s^2}}{2\sqrt{2}m_1} \bigg[ \mathcal{C}_3(-1+z_s) + \mathcal{C}_2(1+z_s) \bigg] \omega_1
 \ , \nonumber \\
V_{0-++}
&=& - \frac{\sqrt{1-z_s^2}}{2\sqrt{2} m_1} \bigg[ \mathcal{C}_2(-1+z_s) + \mathcal{C}_3(1+z_s) \bigg] \omega_1
 \ , \nonumber \\
V_{-0++}
&=& \frac{\sqrt{1-z_s^2}}{2\sqrt{2} m_2} \bigg[\mathcal{C}_2(-1+z_s) + \mathcal{C}_3(1+z_s) \bigg] \omega_2
 \ , \nonumber \\
V_{+0+0}
&=& \frac{(1+z_s)}{2 m_2 m_4} \bigg[ \mathcal{C}_3 (-1+z_s) \omega_2 \omega_4 + \mathcal{C}_2 (-p_{cm} \bar{p}_{cm}  \nonumber  \\
&+& z_s \omega_2 \omega_4 ) \bigg]   \ , \nonumber \\
V_{-0+0}
&=& -\frac{(-1+z_s) }{2m_2 m_4} \bigg[ \mathcal{C}_3 (1+z_s) \omega_2 \omega_4 + \mathcal{C}_2(-p_{cm} \bar{p}_{cm}   \nonumber  \\
&+& z_s \omega_2 \omega_4) \bigg]  \ ,  \nonumber \\
V_{0+0+}
&=& \frac{(1+z_s)}{2 m_1 m_3} \bigg[ \mathcal{C}_3(-1+z_s) \omega_1 \omega_3 + \mathcal{C}_2 (-p_{cm} \bar{p}_{cm}  \nonumber  \\
&+&  z_s \omega_1 \omega_3) \bigg]  \ ,  \nonumber \\
V_{0-0+}
& =&  -\frac{(-1+z_s)}{2 m_1 m_3} \bigg[ \mathcal{C}_3 (1+z_s) \omega_1 \omega_3 + \mathcal{C}_2 (-p_{cm} \bar{p}_{cm} \nonumber  \\
&+& z_s \omega_1 \omega_3) \bigg]  \ , \nonumber  \\
V_{+-+0}
&=& -\frac{1}{2\sqrt{2} m_4} \bigg[(\mathcal{C}_2 + \mathcal{C}_3)(1+z_s) \sqrt{1-z_s^2} \omega_4 \bigg]
 \ , \nonumber \\
V_{0++-}
&=&  -\frac{1}{2\sqrt{2} m_1} \bigg[(\mathcal{C}_2 + \mathcal{C}_3)(-1+z_s) \sqrt{1-z_s^2} \omega_1 \bigg]
 \ ,  \nonumber \\
V_{-++0}
&=& -\frac{1}{2\sqrt{2} m_4} \bigg[(\mathcal{C}_2 + \mathcal{C}_3) (-1+z_s) \sqrt{1-z_s^2} \omega_4 \bigg]
  \ , \nonumber \\
V_{0-+-}
&=& \frac{1}{2\sqrt{2} m_1} \bigg[(  \mathcal{C}_2+ \mathcal{C}_3) (1+z_s) \sqrt{1-z_s^2} \omega_1 \bigg]
 \ , \nonumber \\
V_{0++0}
&=& \frac{(z_s-1)}{2 m_1 m_4} \bigg[ \mathcal{C}_2 (1+z_s) \omega_1 \omega_4\nonumber\\
&&+ \mathcal{C}_3 (p_{cm} \bar{p}_{cm} + z_s \omega_1 \omega_4)\bigg]  \ , \nonumber \\
V_{0-+0}
&=& \frac{(1+z_s)}{2m_1 m_4} \bigg[ \mathcal{C}_2 (-1+z_s) \omega_1 \omega_4 
\nonumber\\
&&
+ \mathcal{C}_3 (p_{cm} \bar{p}_{cm} + z_s \omega_1 \omega_4) \bigg]  \ , \nonumber \\
V_{+-+-}
&=& \frac14 \bigg[ (\mathcal{C}_2+ \mathcal{C}_3) (1+z_s)^2 \bigg]  \ , \nonumber  \\
V_{-++-} &=& 
\frac14 \bigg[ (\mathcal{C}_2 + \mathcal{C}_3) (-1+z_s)^2 \bigg]  \ ,  \nonumber  \\
V_{+-++} &=& 
-\frac14 \bigg[ (\mathcal{C}_2 + \mathcal{C}_3) (-1+z_s^2)  \bigg] \ ,   \nonumber  \\
V_{-+++} &=& 
-\frac14 \bigg[ (\mathcal{C}_2 + \mathcal{C}_3) (-1+z_s^2) \bigg]   \ ,  \nonumber  \\
V_{00+0}
&=& \frac{\sqrt{1-z_s^2}}{\sqrt{2} m_1 m_2 m_4} \bigg[-\mathcal{C}_2 p_{cm} \bar{p}_{cm} \omega_1 + \mathcal{C}_3p_{cm} \bar{p}_{cm} \omega_2 \nonumber  \\
& +&  \mathcal{C}_2 z_s \omega_1 \omega_2 \omega_4 + \mathcal{C}_3 z_s \omega_1 \omega_2 \omega_4 \bigg]  \ ,   \nonumber \\
V_{000+}
&=& - \frac{\sqrt{1-z_s^2}}{\sqrt{2} m_1 m_2 m_3} \bigg[ \mathcal{C}_2 \omega_2(-p_{cm} \bar{p}_{cm} + z_s \omega_1 \omega_3 ) \nonumber  \\
&+& \mathcal{C}_3 \omega_1 (p_{cm} \bar{p}_{cm} + z_s \omega_2 \omega_3) \bigg]  \ , \nonumber
\eea
where $p_{\rm c.m.}=|\vec{p}_1|=|\vec{p}_2|$ and $\bar{p}_{\rm c.m.}=|\vec{p}_3|=|\vec{p}_4|$ are the modulus of the momenta of the initial and final states in the center-of-mass (c.m.) frame, respectively. Furthermore, $\omega_i=\sqrt{m_i^2+|\vec{p}_i|^2}$ ($i=1,\dots,4$) are the energies of the involved states. 
To obtain the above expressions, we have used the following polarization vectors
\bea
&\epsilon_1^\mu(\vec{p}_1,\lambda_1=\pm1) = \dfrac{1}{\sqrt{2}} (0, \mp 1, -i , 0)^T \ ,\notag \\
&\epsilon_1^\mu(\vec{p}_1,\lambda_1=0) = \dfrac{1}{m_1} (|\vec{p}_1|, 0 , 0, \omega_1)^T \ ,  \notag \\
&\epsilon_3^\mu(\vec{p}_3,\lambda_3=\pm1) = \dfrac{1}{\sqrt{2}} (0, \mp \cos \theta, -i , \pm \sin \theta)^T \ ,  \notag \\
&\epsilon_3^\mu(\vec{p}_3,\lambda_3=0) = \dfrac{1}{m_3} (|\vec{p}_3|, \omega_3 \sin\theta , 0, \omega_3 \cos \theta)^T \ ,   
\eea
for particles $1$ and $3$. It is straightforward to obtain the ones for the states $V_2$ and $V_4$ in the c.m. frame. Note that the scattering plane has been chosen to be the plane spanned by the $x$-axis and $z$-axis, such that the azimuthal angle $\phi=0$.

\section{Effective Lagrangian in the heavy quark formalism \label{sec:HQSS}}

\begin{figure*}[tbp]
\begin{center}
\epsfig{file=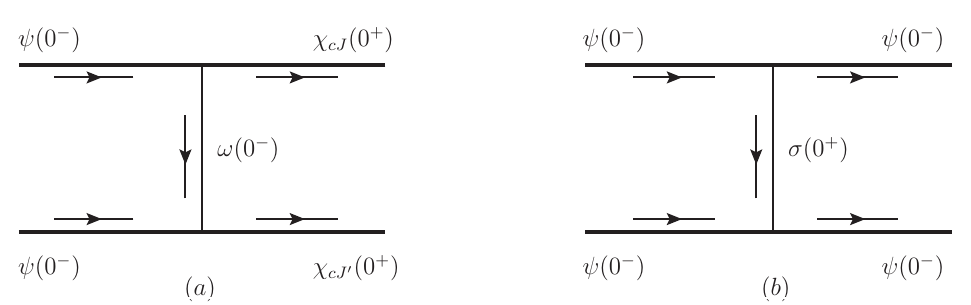,scale=0.6}
\end{center}
\caption{\label{fig:meson_exchange} Feynman diagrams of the two types of scattering processes in meson-exchange picture: (a) $\psi\psi \to \chi_{cJ}\chi_{cJ^\prime}$, (b) $\psi\psi \to\psi\psi$. The quantum numbers $I^{G}$ of the involved particles are shown in the brackets, where $I$ and $G$ are isospin and $G$ parity, respectively.}
\end{figure*}

In the heavy quark formalism, the relevant Lagrangian describing interactions of four charmonium states and respecting heavy quark spin symmetry (HQSS),  can be constructed as
\begin{align}
\mathcal{L}_{\rm HQSS} &= g_1 \langle \bar{\hat{J}}\hat{J} \rangle \langle \bar{\hat{J}}\hat{J}\rangle 
+ g_2\big[ \langle \bar{J}^{\mu} \hat{J} \rangle  \langle \bar{J}_{\mu} \hat{J} \rangle    
+ H.c.\big]\ ,\label{eq:HQSSlag}
\end{align}
where $\langle\cdots\rangle$ represents the trace over the Dirac matrices and $g_1,g_2$ are coupling constants. Here $\hat{J}$ and $J^\mu$ stand for $S$-wave doublet and $P$-wave quartet, respectively. Their explicit expressions read~\cite{ CASALBUONI199395, CASALBUONI1997145, Cincioglu:2016fkm}:
\begin{align}
\hat{J} (L=0) &= \frac{1+ v\!\!\!/ }{2} \bigg[\psi_{\mu}\gamma^{\mu}-\gamma_5 \eta_{c} \bigg]  \frac{1-v\!\!\!/ }{2}   \ , \label{eq:JJ} \\
J^{\mu}  (L=1) &= \frac{1+ v\!\!\!/ }{2}\bigg[\chi_{c2}^{\mu \alpha} \gamma_{\alpha} + \frac{i}{\sqrt{2}}\epsilon^{\mu \alpha \beta \gamma} \chi_{c1 \gamma} v_{\alpha} \gamma_{\beta} \notag \\
&+\frac{1}{\sqrt{3}}\chi_{c0}(\gamma^{\mu}-v^{\mu})  
+h^{\mu}_{c} \gamma_5 \bigg]\frac{1- v\!\!\!/ }{2}  \ ,\label{eq:Jmu}
\end{align}
with $v^\mu$ denoting the four-velocity of the relevant charmonium and $L$ being the orbital angular momentum between the heavy quark and antiquark.
The decomposition of the multiplet for the general case with $L\neq 0$ can be found in Ref.~\cite{CASALBUONI199395}. In the $S$-wave doublet, $\psi_{\mu}$ and $\eta_{c}$ are the vector mesons and pseudoscalar mesons, respectively. For the $P$-wave charmonium bound states, the multiplet $J^{\mu}$ is composed of four states, $\chi_{c2}$, $\chi_{c1}$, $\chi_{c0}$ and $h_{c}$.  The conjugations of the fields are defined by $\bar{\hat{J}}=\gamma^0\hat{J}^\dagger\gamma^0$ and  $\bar{{J}}^\mu=\gamma^0{J}^{\mu\dagger}\gamma^0$. Note that the radial numbers $n$ of the charmonia are suppressed for brevity. 

Now we expand the HQSS Lagrangian by inserting Eqs.~\eqref{eq:JJ} and~\eqref{eq:Jmu} into Eq.~\eqref{eq:HQSSlag}. The $g_1$ term gives
\begin{align}
g_1 \langle \bar{\hat{J}}\hat{J} \rangle \langle \bar{\hat{J}}\hat{J}\rangle 
&= 4g_1 \psi^{\dagger}_ {\mu}  \psi^{\mu} \psi^{\dagger}_ {\nu} \psi^{\nu} 
- 4g_1 \psi^{\dagger}_{\mu}   \psi^{\mu} \eta_{c}^{\dagger} \eta_{c} 
 \notag \\
& 
-4g_1  \eta_{c}^{\dagger} \eta_{c} \psi^{\dagger}_{\nu} \psi^{\nu} 
+4g_1 \eta_{c}^{\dagger} \eta_{c} \eta_{c}^{\dagger} \eta_{c}\ .
\end{align}
In the heavy quark formalism, $\psi_\mu$ ($\eta_c$) annihilates a state,  while its conjugation $\psi_\mu^\dagger$ ($\eta_c^\dagger$) creates a state. Therefore, one can use the $g_1$ term to describe the scattering processes of $\psi\psi\to\psi\psi$, $\psi\eta_c\to\psi\eta_c$, and $\eta_c\eta_c\to \eta_c\eta_c$. However, the process such as $\psi\psi\to\eta_c\eta_c$ does not show up in the Lagrangian, as expected.

The $g_2$ term contains the $\chi_{cJ}$ and $h_c$ states, which can be expanded as
\begin{align}
&g_2\big[\langle \bar{J}^{\mu} \hat{J} \rangle  \langle \bar{J}_{\mu} \hat{J} \rangle  + H.c.\big]
=\big\{4 g_2\chi_{c2}^{\dagger \mu \alpha} \psi_{\alpha} \chi_{c2 {\mu }}^{\dagger \rho} \psi_{\rho} \notag \\
&-2g_2 (\chi^{\dagger}_{c1 \gamma} \psi_{\beta}  \chi^{\dagger {\beta}}_{c1 } \psi^{\gamma}  - \chi^{\dagger}_{c1 \gamma} \psi_{\beta}  \chi^{\dagger {\gamma}}_{c1 } \psi^{\beta} )
+  \frac{4}{3}g_2 \chi^{\dagger}_{c0} \psi^{\mu}  \chi^{\dagger}_{c0} \psi_{\mu}  \notag  \\
&-2 \sqrt{2}g_2 i (\epsilon_{\mu}^{\ \rho \eta \sigma})\chi_{c2}^{\dagger \mu \alpha} \psi_{\alpha} \chi^{\dagger}_{c1 \sigma} v_{\rho} \psi_{\eta} 
+ \frac{4}{\sqrt{3}} g_2\chi_{c2}^{\dagger \mu \alpha} \psi_{\alpha} \chi^{\dagger}_{c0} \psi_{\mu} \notag \\
& - 2\sqrt{2} g_2i (\epsilon^{\mu \alpha \beta \gamma}) \chi^{\dagger}_{c1 \gamma} v_{\alpha} \psi_{\beta} \chi_{c2 {\mu }}^{\dagger \rho} \psi_{\rho} 
 \notag \\
& + \frac{4}{\sqrt{3}} g_2 \chi^{\dagger}_{c0} \psi^{\mu} \chi_{c2 {\mu }}^{\dagger \rho} \psi_{\rho}
- \frac{4}{\sqrt{3}} g_2i (\epsilon_{\mu}^{\ \rho \eta \sigma}) \chi^{\dagger}_{c0} \psi^{\mu}  \chi^{\dagger}_{c1 \sigma} v_{\rho} \psi_{\eta}\notag\\
&- \frac{4}{\sqrt{3}} g_2i  (\epsilon^{\mu \alpha \beta \gamma}) \chi^{\dagger}_{c1 \gamma} v_{\alpha} \psi_{\beta} \chi^{\dagger}_{c0} \psi_{\mu} +{\rm H.c.}\big\}+\cdots\ ,
\end{align}
where \lq\lq$\cdots$'' denotes the pieces with the number of $\psi$ fields less than 2.
It can be seen that the channels of $ \psi \psi \to \chi_{cJ}  \chi_{cJ^\prime} $ ($J,J^\prime=0,1,2$) appear explicitly in the HQSS Lagrangian. On the contrary, the reaction $ \psi \psi\to h_ch_c$ is absent, since it violates HQSS by flipping the charm-quark spin.

We can further estimate the strength of $\psi\psi\to \psi\psi$ and $\psi\psi\to\chi_{cJ}  \chi_{cJ^\prime}$ in the meson-exchange picture. Both of the two types of interactions are OZI allowed. Taking into the SU(3) flavor and isospin symmetries together with $G$-parity conservation into account, the lowest meson can be exchanged for the $\psi\psi\to \psi\psi$ is the $f(500)$ meson (i.e. $\sigma$), while for  $\psi\psi\to\chi_{cJ}  \chi_{cJ^\prime}$ it is the $\omega$ meson, as illustrated in Fig.~\ref{fig:meson_exchange}.\footnote{It should be pointed out that the mechanisms shown in Fig.~\ref{fig:meson_exchange} actually violate OZI rule. Nevertheless, following Ref.~\cite{Dong:2020nwy}, it is assumed that the interaction between the quarkonium states under our consideration is dominated by the exchange of light modes (soft gluons or, e.g., pion pairs). For instance, the exchanged $\sigma$ meson can be effectively regarded as two pions. } By integrating out the exchanged meson, one can expect that 
\bea
g_1\propto \frac{1}{M_\sigma^2}\ ,\qquad g_2\propto \frac{1}{M_\omega^2}\ ,
\eea
where $M_\sigma$ and $M_\omega$ are masses of $f(500)$ and $\omega$, respectively. Since $M_\omega> M_\sigma$, and hence $g_2<g_1$. That is, the coupling of $\psi\psi\to\psi\psi$  is larger than the one of  $\psi\psi\to\chi_{cJ}  \chi_{cJ^\prime}$. 
Therefore, it might be a good approximation to take only the type of process of $\psi\psi\to\psi\psi$ into consideration, when studying the di-$J/\psi$ invariant mass spectrum with limited experimental data.

For convenience, we denote the $4\psi$ interaction in the HQSS Lagrangian by 
\bea
\mathcal{L}_{4\psi}\equiv 4g_1 \psi^{\dagger}_ {\mu}  \psi^{\mu} \psi^{\dagger}_ {\nu} \psi^{\nu}  \subset\mathcal{L}_{\rm HQSS}\ ,
\eea
and briefly discuss its relationship with the effective Lagrangian given in Eq.~\eqref{eq:coneffLag} in Sec.~\ref{sec2}. To that end, the radial quantum numbers of the charmonium states have to be invoked and the $\mathcal{L}_{4\psi}$ Lagrangian is recast into
\bea
\mathcal{L}_{4\psi}&=&4g_1(mn;m^\prime n^\prime)\notag\\
&\times&\psi^{\dagger}_ {\mu}(m^\prime S)  \psi^{\mu}(mS) \psi^{\dagger}_ {\nu}(n^\prime S) \psi^{\nu}(nS)\ .\label{eq:4phi}
\eea
Here $g_1(mn;m^\prime n^\prime)$ denotes the coupling constant corresponding to the process of $\psi(mS)\psi(nS)\to \psi(m^\prime S)\psi(n^\prime S)$. Note that $\psi(1S)$ should be identified as the $J/\psi$ state. Correspondence between the Lagrangian in Eq.~\eqref{eq:4phi} and the one in Eq.~\eqref{eq:coneffLag} can be found. For instance, 
\begin{align}
&4g_1(11;11) \to  h_1\ ,\quad 4g_1(11;12) \to  h_2\ ,\cdots
\end{align}
However, it should be pointed out that the Lagrangian in Eq.~\eqref{eq:coneffLag} is constructed in explicitly relativistic formalism, and all kinds of contraction of Lorentz indices are taken into account. Therefore,  the relativistic Lagrangian has more terms than the HQSS Lagrangian. Taking the interaction of $J/\psi J/\psi\to \psi(2S)\psi(2S)$ for example, 
there are two terms accompanied by $h_4$ and $h_4^\prime$ in the relativistic Lagrangian but only one term in the HQSS one indicated by $4g_1(11;22)$.  The merit of the use of relativistic formalism is that the original analytical properties of the obtained scattering amplitudes are respected.

Finally, it is also worth noting that the incorporation of the $\psi(3700)$ in the HQSS Lagrangian demands the introduction of the $\psi(1D)$ charmonium state. The $\psi(3770)$ state is usually regarded as a mixture of $\psi(1D)$ and $\psi(2S)$ states, and the $1^3D_1$ charmonium component is generally considered to be predominant~\cite{Brambilla:2010cs}. We refer the readers to Ref.~\cite{CASALBUONI199395} for how to introduce the charmonium states with $L=2$. Nevertheless, in our relativistic effective Lagrangian given in Eq.~\eqref{eq:coneffLag}, the $\psi(3700)$ state is directly included as an explicit degree of freedom and its relevant interactions are constrained by symmetries.

\bibliography{X6900refs}

\end{document}